\documentclass[useAMS,usenatbib]{mn2e}

\usepackage{amsmath}
\usepackage{amssymb}

\usepackage{epsfig}
\usepackage{graphicx}

\title{Black Hole/Pulsar Binaries in the Galaxy}

\author[Shao \& Li]{Yong Shao$^{1,2}$\thanks{E-mail:
shaoyong@nju.edu.cn; lixd@nju.edu.cn} and Xiang-Dong Li$^{1,2}$\footnotemark[1]\\
$^{1}$Department of Astronomy, Nanjing University, Nanjing 210046, China\\
$^{2}$Key Laboratory of Modern Astronomy and Astrophysics (Nanjing University), Ministry of
Education, Nanjing 210046, China}

\begin{document}

\date{Accepted 1988 December 15. Received 1988 December 14; in original form 1988 October 11}

\pagerange{\pageref{firstpage}--\pageref{lastpage}} \pubyear{2002}

\maketitle

\label{firstpage}

\begin{abstract}

We have performed population synthesis calculation on the formation of binaries 
containing a black hole (BH) and a neutron star (NS) in the Galactic disk. 
Some of important input parameters,  especially for the treatment of common envelope 
evolution, are updated in the calculation. We have discussed the uncertainties from the star 
formation rate of the Galaxy and the velocity distribution of NS kicks 
on the birthrate ($ \sim 0.6-13 \rm\, Myr^{-1}$) 
of BH/NS binaries. From incident BH/NS binaries, by modelling the orbital evolution 
duo to gravitational wave radiation and 
the NS evolution as radio pulsars, we obtain the distributions of the observable parameters
such as the orbital period, eccentricity and pulse period of the BH/pulsar binaries. 
We estimate that there may be $\sim 3-80  $ BH/pulsar binaries in the Galactic disk and around
10\% of them could be detected by the Five-hundred-meter Aperture Spherical radio Telescope. 

\end{abstract}


\section{Introduction}

More than forty years have passed since the first binary pulsar of PSR B1913+16 was 
discovered \citep{ht75}, and there are now $ \sim 15 $ known in the Galaxy \citep[see][for a compilation]{t17}.
To date, however, no binaries containing a black hole (BH) and a pulsar has been
detected. Discovery of BH/pulsar binaries is a holy grail in astrophysics, not only because they 
can yield constraints on the evolution of massive star binaries, but also they are very useful for 
testing relativistic gravity \citep[e.g.,][]{l14} and understanding the formation of gravitational 
wave sources \citep[e.g.,][]{b02}. There have been investigations on the formation of BH/pulsar binaries 
\citep[][for a review]{s03}, but mainly focusing on binaries containing a recycled pulsar 
\citep{s04,p05} and formed due to dynamical processes \citep{fl11,c14}. For binary systems with a 
recycled pulsar, \citet{p05} concluded that the birthrate in the Galactic disk is probably no
larger than $ \sim 0.1 \,\rm Myr^{-1} $ and even including zero, the expected number is fewer than $ \sim 10 $.
In both the Galactic centre and globular clusters, the estimated number of dynamically formed 
BH/(recycled) pulsar binaries has the order of unity, the upper limit on the size of this population
is $ \sim 10 $ \citep{fl11,c14}.

In this Letter, we revisit the population of BH/pulsar binaries through isolated binary evolution.
Compared to previous studies, we have included new important updates in the population
synthesis calculation to obtain more reliable estimates on the binaries containing a BH and a
neutron star (NS). In addition, 
the NSs in binary systems are regarded as isolated radio pulsars, we can model the pulsar
evolution with time to derive the parameter distribution of BH/pulsar binaries.  
Searching for BH/pulsar systems is one of the important scientific goals of the 
Five-hundred-meter Aperture Spherical radio Telescope (FAST) because of its excellent performance, 
our results can provide helpful information for detectable BH/pulsar binaries in the Galactic disk. 
The rest of the paper is organized as follows. In section 2 we introduce the method in this work. 
We present the results in section 3. A brief discussion is made in section~4.

\section{Method}

\subsection{Population Synthesis Calculation}

To model the population of BH/pulsar systems, we firstly utilize the \textit{BSE} population 
synthesis code \citep{h02} to perform a series of Monte-Carlo simulations for a large number ($ 2\times10^{7} $) 
of binary evolution calculations to obtain the distribution of BH/NS binaries.
Some of the modifications of the code, including the process of mass transfer,
the supernova explosions and the treatment of common envelope (CE) evolution, 
have been described by \citet{sl14}.
The key points are summarized as follows.

In a primordial binary, the first mass transfer occurs when the primary star evolves to 
overflow its Roche lobe. If the mass
transfer proceeds very rapidly, the secondary star will get out of thermal equilibrium due to mass
accretion, and the responding expansion may cause the secondary star to fill its own Roche lobe,
leading to the formation of a contact binary \citep{ne01}. In the calculation we assume that
contact binaries will go into CE evolution. The mass transfer efficiency (i.e. the fraction of
mass accreted onto the secondary star among the transferred mass) is an important factor that determines whether a
binary will evolve to be contact. Here we assume the efficiency to be 0.5, based on \citet{sl14} for the
formation of Galactic Be/X-ray binaries. The corresponding critical mass ratio between the primary
and the secondary stars for stable mass transfer is 2.0 \citep{sl14}.

For dynamically unstable mass transfer in a binary, we use the standard energy conservation 
equation \citep{w84} to deal with the CE evolution. During the spiral-in stage, the orbital 
energy of the accretor is used to balance the binding energy of the donor envelope. We follow
\citet{xl10} and \citet{w16} to deal with the binding energy parameter $ \lambda $ in which the contribution of 
the internal energy is included, and the CE efficiency $ \alpha_{\rm CE} $ is taken to be 1.0 
in the calculation.

A massive star will finally evolve to produce a compact star. We set the remnant mass
according to the rapid supernova mechanism \citep{f12}, which can account for the $ \sim2-5M_{\odot} $
gap or deficiency between the measured masses of NSs and BHs. 
At the moment of supernova explosions, the compact stars are subject to supernova kicks.
\citet{h05} proposed that the distribution of the pulsar birth velocities is well fitted by 
a Maxwellian distribution with $ \sigma = 265 \rm \,km\,s^{-1} $, while \citet{v17} argued that
a bimodal distribution (with $ \sigma \sim  80 $ and $ 320 \rm \,km\,s^{-1} $) can better describe 
the birth velocities, among which the lower distribution may be attributed to the pulsars 
formed from electron-capture supernovae. Note that the birth velocities include the contribution
from supernova kicks, the pre-supernova binary motion and the motion of the progenitor systems. In our
calculation, the NS kick velocities are assumed to have a Maxwellian distribution, we discuss three
different velocity distributions with $ \sigma_{\rm k} = 50$, 150 and  $300 \,\rm \,km\,s^{-1} $.
For natal kicks to newborn BHs, we use the NS kick velocities reduced by a factor of $ (1-f_{\rm fb} )$, 
where $ f_{\rm fb} $ is the fallback fraction.

The general scenario for the formation of BH/NS binaries is that a massive BH forms first, with a 
NS at a later epoch \footnote{An alternative scenario is that the BH progenitor transfers
a great amount of material to the NS progenitor, resulting in the NS being formed first. 
The NS is then recycled by the transferred matter from the BH progenitor. 
This requires that enough mass must be transferred.
From our calculation, there is no such binaries produced since the adopted mass transfer 
efficiency is 0.5. Furthermore, binary population synthesis studies demonstrated that these 
systems are very rare and probably not existent \citep{s04,p05}.}. During the evolution 
in binaries with a BH accretor and a massive normal donor (the NS progenitor), we 
use the \textit{EV} stellar evolution code initially developed by \citet{e71} to deal with the 
mass transfer stability. The CE evolution is assumed to be triggered when either the donor 
has developed a deep convective envelope
(with the mass transfer rate rapidly reaches $ \gtrsim 0.01 \,M_{\odot}\,\rm yr^{-1} $) or the trapping 
radius \footnote{For binaries with a BH accretor, there is a so-called “trapping radius” in the 
super-Eddington accretion flow around the BH, below which photon diffusion outward cannot overcome 
the advection of photons inward, and the radiation generated in excess of the Eddington limit can thus
be advected into the BH \citep{ss73}. \citet{kb99} proposed
that CE evolution will be avoided if the trapping radius is smaller than the BH's Roche lobe
radius.} of the accreting BH is larger than its Roche lobe radius. Fig.~1 shows the parameter
spaces of the binaries with the occurrence of CE evolution at the metallicity of 0.02. The black, 
red and green curves correspond to the  
the initial BH mass of 5, 10 and $20M_{\odot}$, respectively. The dashed vertical lines indicate the 
donor's mass of $ 100M_{\odot}$. The maximal mass ratio for avoiding CE evolution can reach as high as
$ \sim 6 $, and the mass transfer may be always stable in some binaries with long periods. This result
is roughly consistent with that of \citet{p17}. In the \textit{BSE} code, we use the interpolation of 
mass ratios and orbital periods (rather than a constant value) to decide whether the CE evolution 
is triggered in binaries with different BH masses.

\begin{figure}
\begin{center}
\includegraphics[scale=0.5]{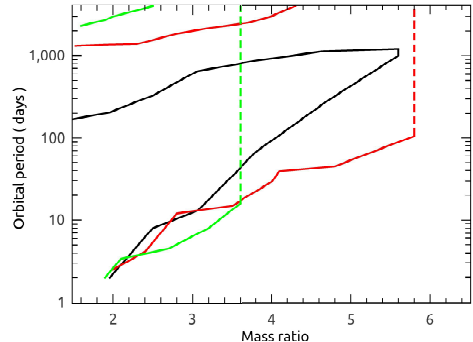}
\caption{The parameter spaces for stable and unstable mass transfer in binaries with  an accreting BH 
in the mass ratio vs. orbital period plane. Binaries located outside the confined region will experience 
unstable mass transfer that leads to CE evolution. The black, red and green curves correspond to the 
initial BH mass of 5, 10 and $20M_{\odot}$, respectively. The dashed vertical lines indicate that the upper
limit ($ 100M_{\odot} $) of the initial donor mass considered in this work. 
 \label{figure2}}
\end{center}
\end{figure}

In our calculation, the initial input parameters of the primordial binaries are set as follows. We adopt the 
\citet{k93} initial mass function for the primary mass and a flat distribution between 0 and 1 for 
the mass ratio of the secondary to the primary \citep{sd12}. All stars are assumed to initially in binaries.
The distribution of the orbital separations 
is assumed to be flat in logarithm and all binaries are assumed to have circular orbits \citep{h02}. 
The primordial binaries are believed to follow the distribution of stars in the Galactic disk and 
the progenitors of BH/NS binaries are assumed to be close to their birth locations (because of the
short lifetime of massive stars and the small kick velocity at BH formation). 
We only track the spatial motions of BH/NS binaries in the Galaxy, which are determined by the Galactic 
gravitational potential \citep{kg89}.
We set the initial metallicity of stars to be 0.02, and assume a constant star formation rate $ S $ 
over a 10 Gyr period. 
Several investigations estimated that the star formation rate of the Galaxy can vary in a range 
of $ \sim 1-5 M_{\odot}\,\rm yr^{-1}$ \citep[e.g.,][]{s78,d06,rw10}. So  
we consider three different values of $ S = 1$ , 3 and
$ 5 M_{\odot}\,\rm yr^{-1}$ for comparison. We will see that the expected 
birthrate of BH/NS binaries vary by several times when adopting different 
star formation rates, but the overall shapes of the parameter 
distribution remain unchanged.

\subsection{Modelling Pulsar Evolution}

\begin{table}
\begin{center}
\caption{The expected number of detectable BH/pulsar binaries by considering different
star formation rates $ S $ and NS kick velocity distributions. Here $ R_{\rm birth} $ denotes 
the birthrate of BH/NS binaries in the Galactic disk, $N_{T}$ the total number
of detectable BH/pulsar binaries, $N_{PM}$ and $N_{FAST}$ the predicted 
number that can be detected in the Parkes Multibeam and FAST surveys with the flux density
limit of $ \rm 0.2\,mJy $ \citep{m01} and $\rm 0.005\,mJy $ (D. Li, private communication)
in the telescope's visible sky, respectively.\label{tbl-1}}
\begin{tabular}{ccccccllll}
\\
\hline
 $ S $ &  $  \sigma_{\rm k}$ & $ R_{\rm birth} $ & $N_{T}$ &$N_{PM}\,^{*}$ & $N_{FAST}\,^{*}$   \\
 ($M_{\odot}\,\rm yr^{-1}$) & ($ \rm km\,s^{-1} $) & ($ \rm Myr^{-1} $) &  & &    \\
\hline
1 & 50& 2.7 & 16& 0.2 & 1.5 \\
 & 150& 1.3 &8 & 0.1 & 0.9 \\
 & 300 & 0.6 & 3 &0.04 & 0.3  \\
3 & 50 & 8.0& 48 &0.5& 4.5  \\
 & 150 & 4.0 & 24 & 0.3 & 2.7 \\
 & 300 & 1.8 & 10 & 0.1 & 0.9  \\
5 & 50 & 13.3 & 80 & 0.8& 7.5  \\
 & 150 & 6.7 & 40 & 0.5& 4.5  \\
 & 300 & 3.0 & 17 & 0.2& 1.5 \\
\hline
\end{tabular}
\end{center}
* In some situations, the calculated number may be smaller than one, it
indicates how low is the probability of one system to be detected.
\end{table}

\begin{figure}
\begin{center}

\includegraphics[scale=0.4]{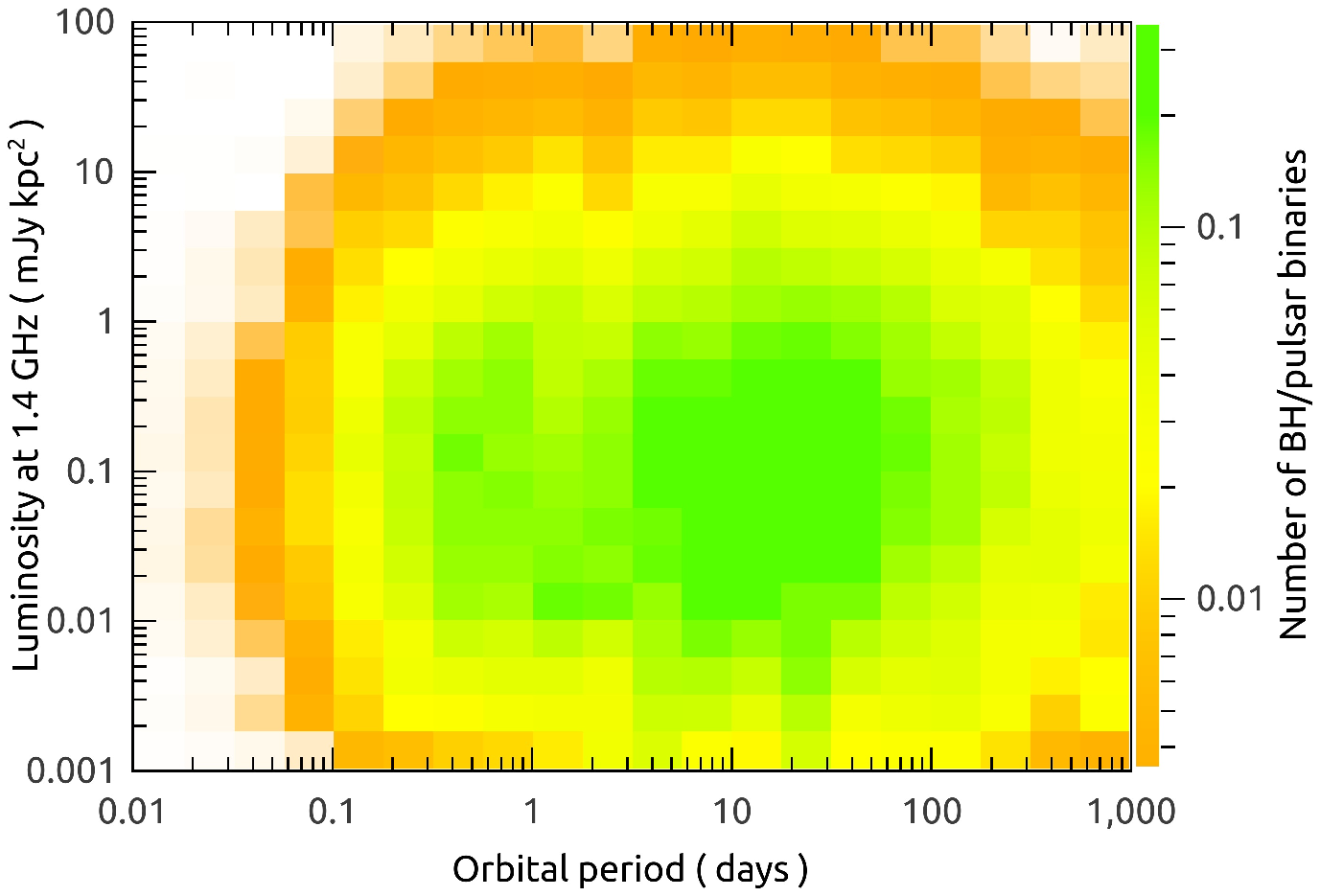}
\includegraphics[scale=0.4]{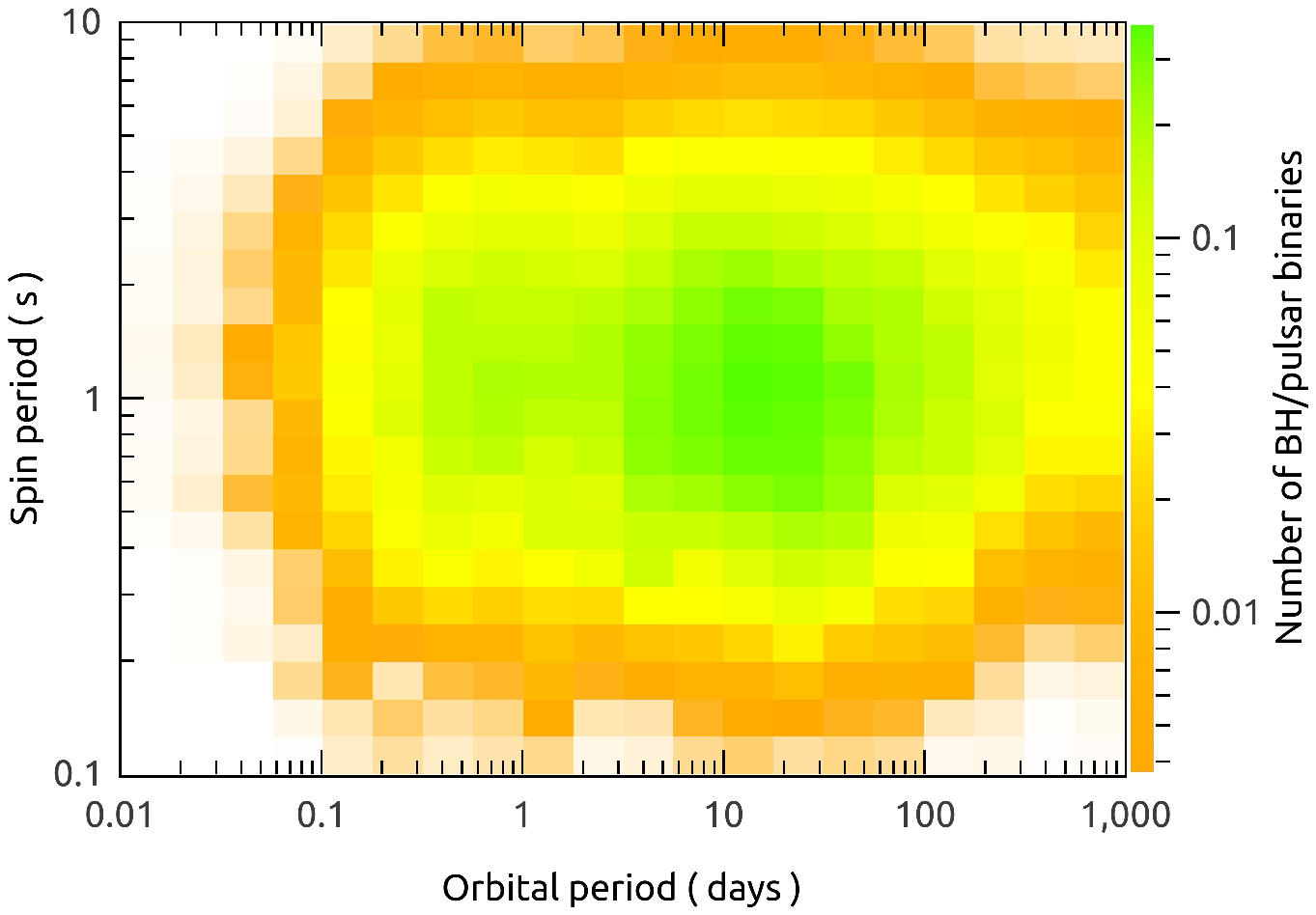}
\caption{The expected distribution of BH/pulsar binaries in the orbital period vs. radio
luminosity (top) and spin period (bottom) planes. 
Here we take the star formation rate to be $ 3 \,M_{\odot}\,\rm yr^{-1}$ and the 
NS kick velocity distribution with $ \sigma_{\rm k} =150 \,\rm km\,s^{-1} $.
\label{figure6}}
\end{center}

\end{figure}

After the formation of a BH/NS binary,
the subsequent orbital evolution is controlled by gravitational wave radiation \citep{p64}. 
For every newborn NS in binaries, we follow the method
of \citet[][and references therein]{fk06} to deal with its evolution as an isolated radio pulsar.
The radio luminosity $ L $ is assumed to
have a power-law dependence on the spin (or pulse) period $ P $ and its period derivative $ \dot{P} $.
Since most pulsars were detected
at 1.4 GHz in previous surveys, we only consider the luminosity at that frequency.
The initial spin period and surface magnetic field strength of the pulsars are assumed to  
obey the log-normal distributions, then we can model their spin evolution that is driven by magnetic 
dipole radiation. The radio emission is assumed to turn off when the pulsars evolve across the
``death line" in the $ P-\dot{P} $ diagram, so we use the relation given by \citet{b92} to classify 
the pulsars as either active or extinguished radio sources.
In addition, we assume that detection of pulsars is constrained by the beaming effect, for which the fraction
of pulsars that beam toward us depends on the spin period $ P $ \citep{tm98}.

\section{Results}

\begin{figure}
\begin{center}

\includegraphics[scale=0.4]{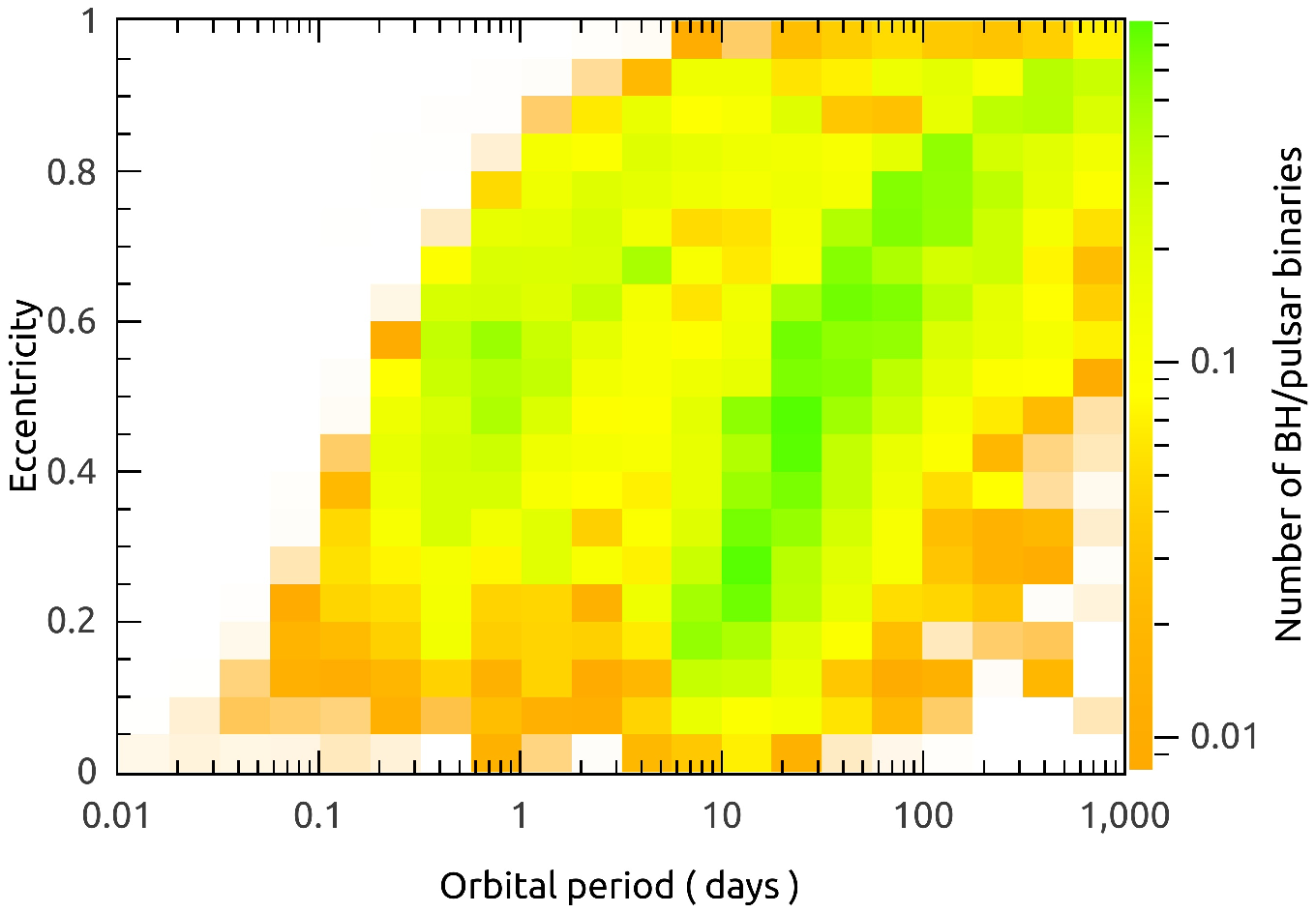}
\includegraphics[scale=0.4]{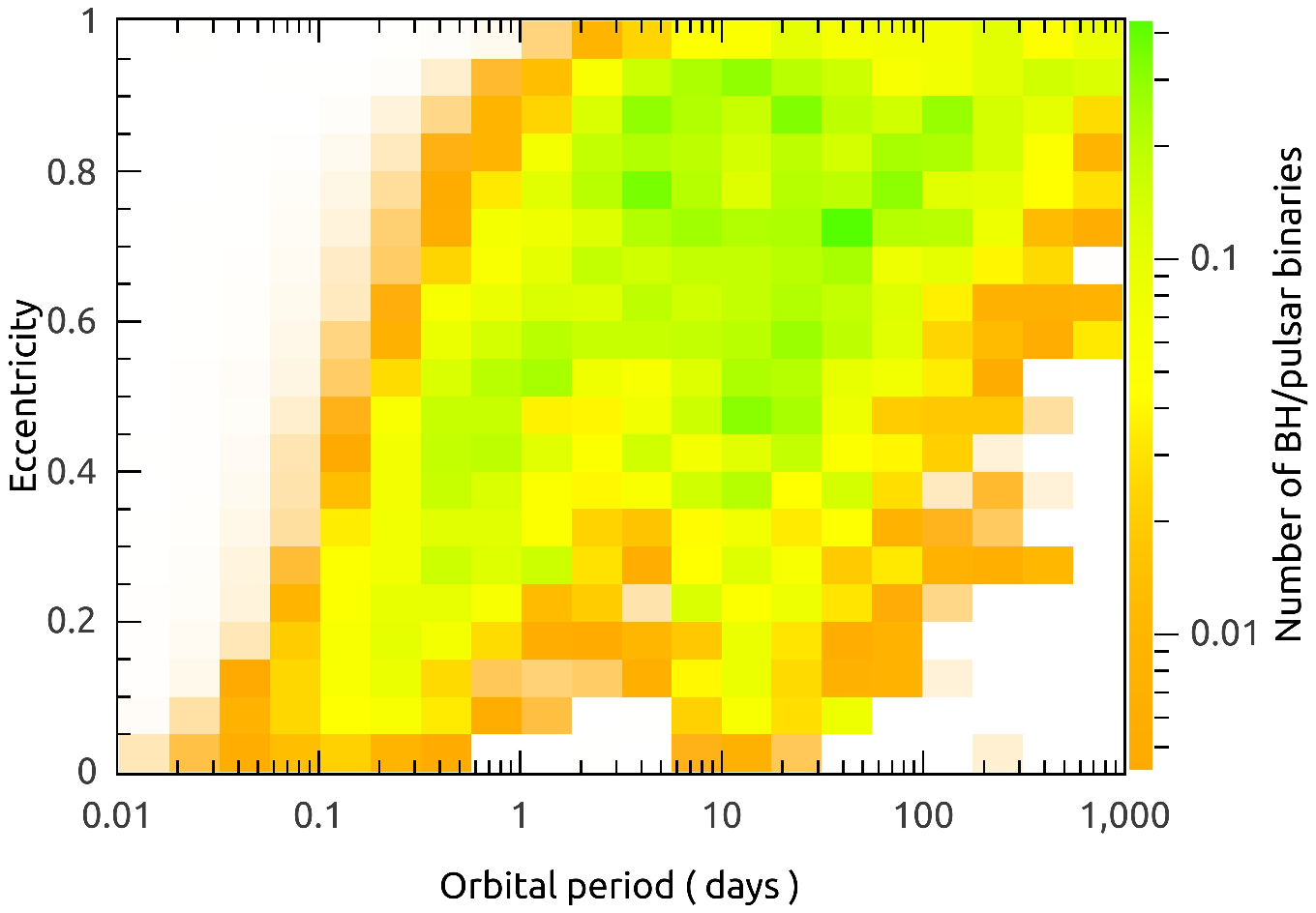}
\includegraphics[scale=0.4]{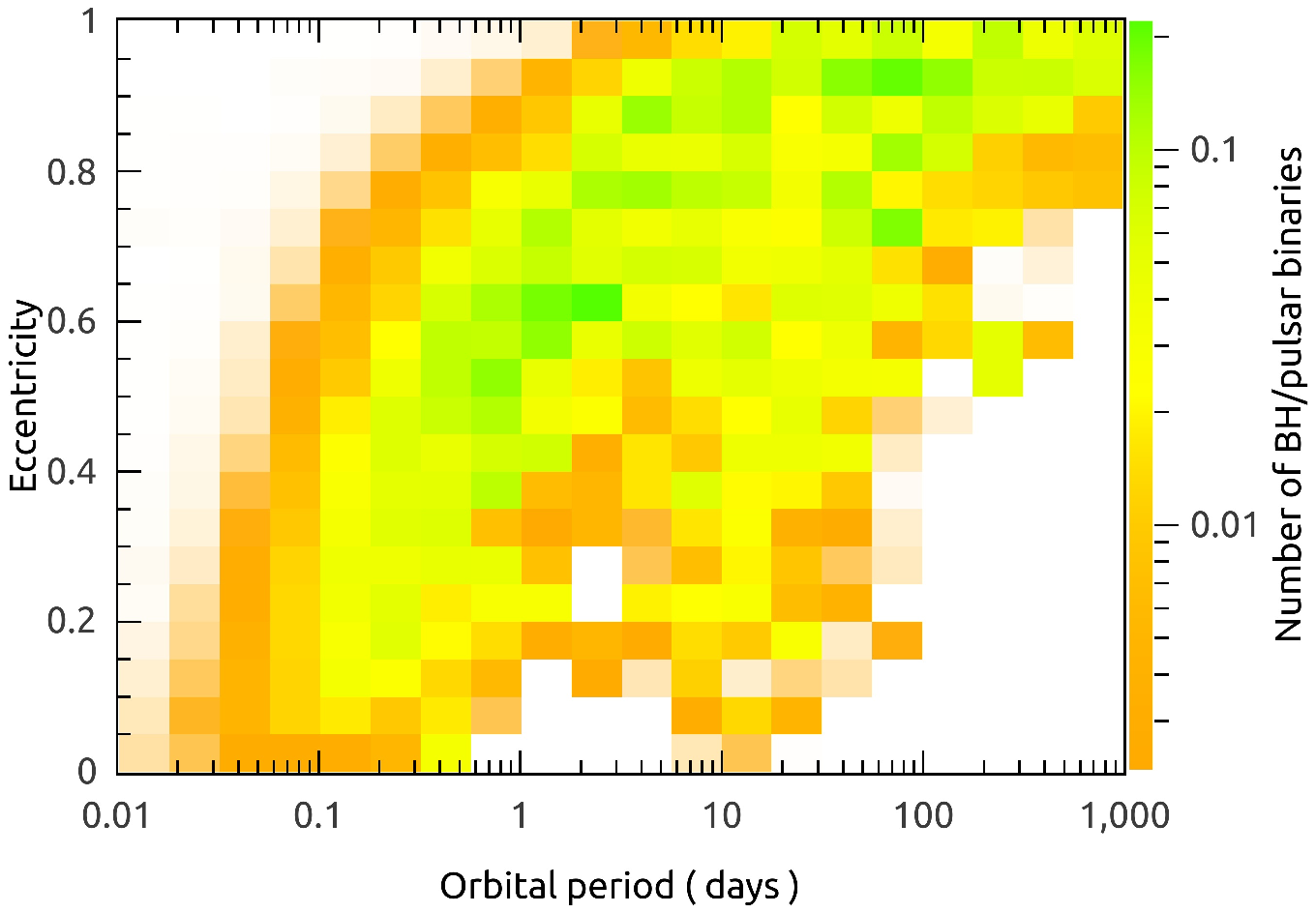}
\caption{The distribution of BH/pulsar binaries in the plane of orbital period and eccentricity.
The top, middle and bottom panels correspond to the NS kick velocity distribution with 
$ \sigma_{\rm k} = 50$, 150 and   $300 \,\rm km\,s^{-1} $, respectively. In each panel, the star
formation rate of the Galaxy is set to be $ 3\,M_{\odot}\,\rm yr^{-1}$.
\label{figure6}}
\end{center}

\end{figure}




For each BH/NS binary produced in our population synthesis calculation, we can obtain the evolutionary 
tracks of the radio luminosity $ L $ and the spin period $ P $ against the pulsar age $ t $.
The number of BH/pulsar binaries in each  interval of $ \Delta L $ and $ \Delta P $ can be evaluated by
accumulating the product of their birthrate and the evolution time $ \Delta t $.
In Table 1, we present the calculated numbers of detectable BH/pulsar binaries considering different
star formation rates and NS kick velocity distributions.
The birthrate of all BH/NS binaries in the Galactic disk varies in the range of 
$ \sim 0.6-13 \rm\, Myr^{-1}$, and
the inferred number (over a period of 10 Gyr) is of the order $ 10^{4} $.
About 3-80 among them can be detected as BH/pulsar binaries.
In the top panel of Fig.~2, by taking $ S = 3 \,M_{\odot}\,\rm yr^{-1}$ and  
$ \sigma_{\rm k} = 150\,\rm km\,s^{-1} $ for an example, we present the distribution (with a total number of 24)  
of BH/pulsar binaries in the plane of orbital period and pulsar luminosity.
The BH/pulsar systems are distributed in a wide range of the orbital
period $\sim  0.1-1000 $ days. The logarithmic luminosity of the pulsars roughly follows a normal 
distribution with the peak of $  0.07 \,\rm mJy\,kpc^{2} $, which is in agreement with the 
results of \citet{fk06}. In the bottom panel of Fig.~2, we plot the distribution of BH/pulsar 
binaries in the plane of orbital period and pulse period. The pulse periods are mainly distributed 
in the range of $\sim 0.4-2 \rm\,s$.

The distribution of BH/pulsar binaries 
in the plane of orbital period and eccentricity is given in Fig.~3. The top,
middle and bottom panels correspond to $ \sigma_{\rm k} =50$, 150 and 
$ 300\, \rm km\,s^{-1} $, respectively. In the case with a low $ \sigma_{\rm k}$ of 
$ 50\, \rm km\,s^{-1} $, the orbital period 
distribution is separated into two groups by the period of several days, which 
reflects whether the progenitor binary with a BH accretor has experienced CE evolution. 
When increasing the kick velocities, more binaries (especially the
long period systems) will be disrupted, and the survived BH/pulsar binaries tend 
to have large eccentricities.

Fig.~4 shows the flux density distribution for the pulsars in BH/pulsar binaries when assuming 
the NS kick velocity distribution with $ \sigma_{\rm k} = 50$ (top), 150 (middle) and  $300 \,\rm km\,s^{-1} $ 
(bottom). The black, red and green curves correspond to
the star formation rate of 1, 3  and $ 5 \,M_{\odot}\,\rm yr^{-1}$, respectively.
It can be seen that the number of BH/pulsar binaries with radio flux density large than
$ 10^{-5} \rm mJy$ is in the range of $ \sim 2-50 $. 
In Table 1, we also give the predicted number of BH/pulsar binaries that can be detected in the 
Parkes Multibeam and FAST surveys with the flux density
limit of $ \rm 0.2\,mJy $ \citep{m01} and $\rm 0.005\,mJy $ (D. Li, private communication).
From the Parkes Multibeam survey, the detected number is expected to be less than 0.8.
This small number may explain the lack of such binaries by now.
Considering the fact the FAST survey will cover around 44\% of the Galaxy \citep{lp16}, we expect 
that the detected number of BH/pulsar binaries in FAST survey is about $\sim 0.3-8 $,
the detection of such binaries is promising.

\begin{figure}
\begin{center}
\includegraphics[scale=0.4]{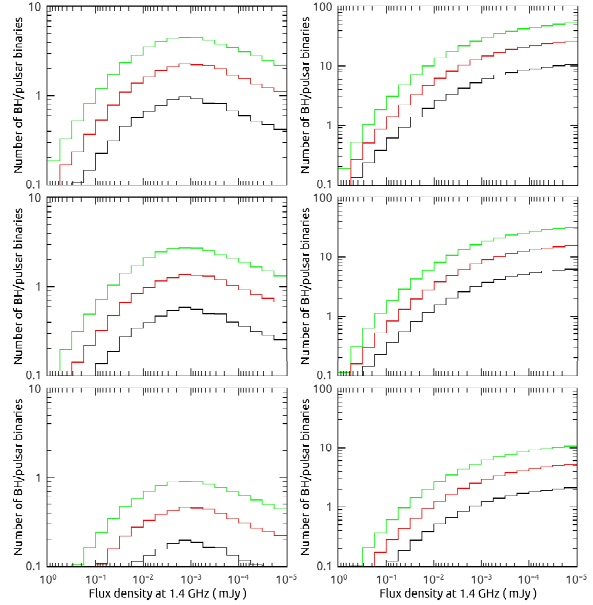}
\caption{The predicted number distribution of BH/pulsar binaries as a function of
the pulsar flux density, when adopting the NS kick velocity distribution with 
$ \sigma_{\rm k} = 50$ (top), 150 (middle) and  $300 \,\rm km\,s^{-1} $ (bottom). 
The black, red and green curves correspond to
the star formation rate of 1, 3  and $ 5 \,M_{\odot}\,\rm yr^{-1}$, 
respectively. The left and right panels denote the differential and cumulative distributions.
\label{figure6}}
\end{center}

\end{figure}


\section{Discussion}

We briefly discuss possible factors that may influence our results. The kick velocity distributions 
(with $ \sigma_{\rm k} = 50 - 300 \,\rm km\,s^{-1} $) at NS birth can change the number 
of BH/pulsar binaries by a factor of $ \sim 4 $. The lower kick velocity, the more BH/PSR binaries can be produced.
When combining the uncertainty in the star formation rate ($ 1-5 \,M_{\odot}\,\rm yr^{-1}$) of the Galaxy, 
the excepted number of detectable BH/pulsar binaries can vary by a factor of $ \sim 20 $. 
In our calculations, we assume that all stars are initially in binary systems. Observations show that the binary 
fraction among all massive stars is $ \sim 70\% $ \citep{sd12}, this can slightly reduce the size of the
BH/pulsar population. During the progenitor evolution of BH/NS binaries, we have accepted some typical 
assumptions. For the process of CE evolution, if we reasonably decrease the CE efficiency $ \alpha_{\rm CE} $ 
from 1.0 to be 0.5, the birthrate of BH/NS binaries will drop by a factor of $ \sim 2 $. 

For spin evolution of the pulsars, we assume the standard model 
of magnetic dipole radiation. The measured braking indices of young pulsars vary in a 
range of $ \sim 0.9-3 $ \citep[][and references therein]{a16}. By simulating the pulsar
evolution, with a normal distribution of braking indices with mean 2.4 and standard deviation 
0.6, \citet{fk06} demonstrated that the corresponding $ P-\dot{P} $ diagrams were found to be 
qualitatively very similar to the one obtained with all braking indices set to 3.
So that the model of magnetic dipole radiation should be a very good approximation.

There is a possibility that a BH/pulsar binary has been existed in the detected pulsar sample but
just not recognized. In the Galaxy, the supernova rate is about $ 0.02 \,\rm yr^{-1} $,
and the number of detected isolated pulsars is  $ \sim 2000 $ \citep{m05}. Based on the formation rate of
$ \sim 0.6-13 \rm\, Myr^{-1}$ for BH/NS binaries from our calculation,
we can estimate that the number of potential BH/pulsar systems is $ \lesssim 1.3 $, which is consistent 
with the calculated upper limit of 0.8 in Parkes Multibeam survey (see Table 1). 
The FAST prediction for BH/pulsar binaries is subject to many uncertainties, the high kick velocity 
and the low CE efficiency are likely to bring a null result in the whole survey.


\section*{Acknowledgments}

We thank the referee Duncan Lorimer for constructive suggestions that improved this Letter
and Prof. D. Li for helpful conversation.
This work was supported by the Natural Science Foundation of China (Nos. 11603010,
11133001, 11333004, 11773015, and 11563003), the National Program on Key Research and Development
Project (Grant No. 2016YFA0400803), and the Natural Science Foundation for the Youth of
Jiangsu Province (No. BK20160611).


\clearpage

\end{document}